\documentclass[pre,twocolumn,showpacs,preprintnumbers,amsmath,amssymb,superscriptaddress]{revtex4}
\usepackage{epsfig}
\usepackage{subfigure}
\usepackage{graphicx}
\usepackage{dcolumn}
\usepackage{bm}
\usepackage{epsfig}

\begin{document}

\preprint{APS/123-QED}

\title{Numerical Study of Wave Propagation
in Uniaxially Anisotropic Lorentzian Backward Wave Slabs}

\author{M.K. K\"arkk\"ainen}
\affiliation{Radio Laboratory, Helsinki University of Technology,
P.O. Box 3000, FIN-02015 HUT, Finland} \email{mkk@cc.hut.fi}

\date{\today}

\begin{abstract}
The propagation and refraction of a cylindrical wave created by a
line current through a slab of backward wave medium, also called
left-handed medium, is numerically studied with FDTD. The slab is
assumed to be uniaxially anisotropic. Several sets of constitutive
parameters are considered and comparisons with theoretical results
are made. Electric field distributions are studied inside and
behind the slab. It is found that the shape of the wavefronts and
the regions of real and complex wave vectors are in agreement with
theoretical results.
\end{abstract}





\maketitle

\section{Introduction}

Metamaterials have received much attention during the last years,
because they possess unusual electromagnetic properties, like, for
example, the opposite directions of phase and group velocities.
Double negative (DNG) materials have negative permittivity and
permeability, and they belong to the class of metamaterials. These
media that are capable of supporting backward waves, have been
also called backward wave (BW) media in the literature
\cite{Lindell}. In BW media, the refraction phenomenon is
anomalous in the sense that the power flow is refracted
negatively, i.e. to the same side of the normal of the interface.
As discussed in \cite{Lindell}, it is not necessary for the medium
to be a DNG medium to be able to support backward waves, because
anomalous refraction can also be realized with anisotropic media
with only one negative material parameter. We will use the term BW
medium throughout the rest of the paper.

The pioneering work on BW materials by Veselago \cite{Veselago},
where slab lenses were mentioned, has gained much attention during
recent years, despite some differing opinions on the subject
\cite{Valanju}. Isotropic BW materials are often called Veselago
materials. The possibility to realize a perfect lens with
isotropic BW slabs was discussed by Pendry in \cite{Pendry}.
Numerical and theoretical considerations of wave propagation in
isotropic BW slabs excited with a line current above the slab were
presented by Ziolkowski and Heyman in \cite{Ziolkowski}. Guidance
of waves in a slab of uniaxially anisotropic metamaterial has been
theoretically discussed by Lindell and Ilvonen in \cite{Lindell2}.

Controversial opinions regarding negative refraction effect and
the perfect lens call for more detailed studies of the wave phenomena
in backward-wave media.
In this paper, wave propagation through uniaxially anisotropic BW
slabs is numerically studied, and comparison is made with the
theory presented in \cite{Lindell}. The theory predicts that there
are regions in certain BW media, where the wave vector becomes
complex, thus resulting in exponentially decaying waves. These
regions are bounded by the asymptotes of the wave vector surfaces,
which can be shown to be hyperbola. We study these phenomena
numerically using the finite-difference time-domain (FDTD) method
in a 2D-problem of a line current radiating in the vicinity of a
BW slab. Also, the existence of surface waves on the interface
between free space and BW medium is demonstrated with an example
case. The BW medium is realized with Lorentzian constitutive
parameters having a single pole pair.

An example problem with some theoretical discussion are presented
in section $3$. Results from the numerical simulations are shown
and discussed in section $4$. Our numerical simulations show that
the wave propagation and refraction phenomena heavily depend on
the parameter choices of the BW medium and are qualitatively in
agreement with the theory.

\section{Numerical Model of Dispersive Medium}

The constitutive relations for a frequency dispersive isotropic
medium read
\begin{equation}
{\mathbf D} = \epsilon(\omega) {\mathbf E},\quad {\mathbf B} =
\mu(\omega) {\mathbf H}. \label{eq:conrel}
\end{equation}
Negative permittivity and permeability are realized with the
Lorentz medium model. The expressions for the permittivity and
permeability are of the form
\begin{eqnarray}
\epsilon(\omega) & = & \epsilon_0 \left( 1+
\frac{\omega_{pe}^2}{\omega_{0e}^2-\omega^2+j\Gamma_e \omega}
\right), \nonumber \\
\mu(\omega) & = & \mu_0 \left( 1+
\frac{\omega_{pm}^2}{\omega_{0m}^2-\omega^2+j\Gamma_m \omega}
\right). \label{eq:perms}
\end{eqnarray}
This model corresponds to a realization of BW materials as
mixtures of conductive spirals or omega particles, as discussed in
\cite{Tretyakov}. In this artificial material both electric and
magnetic polarizations are due to currents induced on particles of
only one shape, which provides a possibility to realize the same
dispersion rule for both material parameters, as in
(\ref{eq:perms}). Note that the medium realized by Smith {\it et.\
al.} is built using different principles \cite{Smith}. For the
uniaxial materials that we consider in this paper we assume that
the negative components of the material parameters are realized by
small uniaxial spiral inclusions (racemic arrays with equal number
of right- and left-handed particles) and possess frequency
dispersion defined by (\ref{eq:perms}). The positive components of
the material parameters are equal to the free-space permittivity
and permeability values, assuming that there are no particles
oriented along these axes.

Equations~(\ref{eq:conrel}) and (\ref{eq:perms}) form the basis of
the used FDTD model for BW materials. The two most important known
FDTD methods for modeling dispersive materials like Lorentz
materials are the recursive convolution method and the auxiliary
differential equation method, where the constitutive parameters
are expressed with the help of susceptibility. In the first
method, ${\mathbf D}$ and ${\mathbf E}$ and ${\mathbf B}$ and
${\mathbf H}$ are related through a convolution integral. This
approach is rather tedious. Another possibility is to use the
auxiliary differential equation technique, which is slightly
easier to implement. In this latter method, the polarization
current associated to each Lorentz pole pair is introduced. These
two models are discussed in detail in \cite{Taflove}.

A third method to discretize fields in Lorentz medium, classified
as direct integration method in \cite{Young}, is based on the
direct discretization of the PDE representing the time-domain
equivalent of the simplified frequency-domain constitutive
relation. The proposed discretization scheme is a modification of
this method. The idea is to transform (\ref{eq:conrel}) into the
time domain using the relation $j\omega \leftrightarrow
\partial/\partial t$ with one integration before discretization
using center differences. Usually, FDTD models based on the
constitutive relation are directly (after multiplication with the
denominator) discretized, as discussed in a summary of FDTD
algorithms for dispersive media in \cite{Young}. We found in
\cite{motl2} that one integration prior to discretization leads to
much better accuracy and, what is also important, to considerably
better stability properties than the direct discretization without
integration. In the numerical simulations, we have used the model
which is discussed in detail in \cite{motl2}.

\section{An Example Problem and Theoretical Discussion}

\begin{figure}[htb]
\centering \epsfig{figure=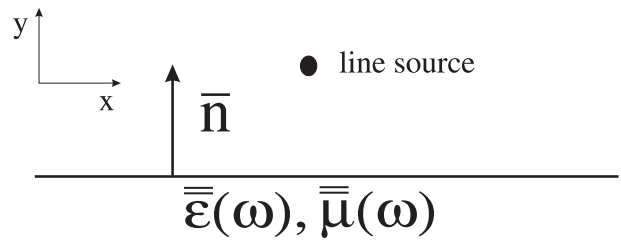, width=0.45\textwidth}
\caption{The slab problem under consideration and the chosen
coordinate system.} \label{kuva0}
\end{figure}

Consider a $z$-directed line current in free space located at a
distance $d_s$ from a BW slab of thickness $d$. Let the interface
between free space and the BW slab be located at $y=0$. The
problem space is two-dimensional, with the field components $H_x,
H_y$, and $E_z$. The peak of the incident spectrum is at
$\omega_{p}=5.0 \cdot 10^9$ rad/s, and the parameters in
(\ref{eq:perms}) are the following: $\omega_{0e}=\omega_{0m}=1.0
\cdot 10^9$ rad/s, $\omega_{pe}^2=\omega_{pm}^2=4.8 \cdot 10^{19}$
(rad/s$)^2$, $\Gamma_e=\Gamma_m=0$. With these choices, we obtain
$\epsilon(\omega)=\mu(\omega)$ for all $\omega$ and
$\epsilon(\omega)/\epsilon_0=\mu(\omega)/\mu_0=-1$ at
$\omega=\omega_{p}$. The spatial resolution $\Delta x=\Delta
y=1.5$ cm is used throughout the simulations. To be able to
demonstrate the properties of BW materials, the incident spectrum
is quite narrow, so that the relative constitutive parameters are
close to minus one for the frequencies having significant spectral
content. Absorbing boundary conditions are used to terminate the
computational domain at the outer boundaries of the lattice. For
simplicity, we have used Liao's third order ABC, although more
sophisticated ABC's are available. The use of usual ABC's requires
a small gap between the outer boundary of the computational space
and the BW material slab. The chosen coordinate axes and the
problem geometry is shown in Figure \ref{kuva0}.
Next, we briefly present some theoretical results from
\cite{Lindell} that are important here for comparison purposes
with our numerical results. Wave propagation in BW slabs is
studied with different value combinations of the medium parameters
$\mu_x$, $\mu_y$ and $\epsilon_z$. For our TE-polarized case, the
Poynting vector ${\mathbf S}_{TE}$ can be shown to read
\cite{Lindell}
\begin{equation}
{\mathbf S}_{TE} = \frac{{\overline {\overline \mu}} \cdot
{\mathbf k}_{TE}}{2 k_0 \eta_0 \mu_0 \mu_x \mu_y} |E_0|^2,
\label{eq:poynt}
\end{equation}
where $|E_0|$ is the amplitude of the TE-polarized electric field
of the plane wave, ${\mathbf k}_{TE}$ is the wave vector with two
cartesian components $k_x$ and $k_y$,
$k_0=\omega\sqrt{\epsilon_0\mu_0}$ is the free space wave number,
and $\eta_0=\sqrt{\mu_0/\epsilon_0}$ is the free space wave
impedance. Denoting the angle between the outwards-pointing unit
normal vector ${\mathbf u}_y$ and the wave vector ${\mathbf
k}_{TE}$ by $\theta$, one can derive the dispersion equation and
solve it for the wave number as in \cite{Lindell}. The result is
\begin{equation}
k_{TE}(\theta) = k_0 \sqrt{\frac{\mu_x
\epsilon_z}{\cos^2\theta+\frac{\mu_x}{\mu_y} \sin^2\theta}}.
\label{eq:ksurf}
\end{equation}
For any given set of parameters, we may plot the projection curves
of the wave vector surfaces in $xy$-plane. For certain choices of
the medium parameters, the wave number becomes complex, resulting
in exponentially decaying waves inside the BW material.

Any physical power flow must be directed downwards away from the
source. This requires that
\begin{equation}
{\mathbf u}_y \cdot {\mathbf S}_{TE} < 0. \label{eq:physcond}
\end{equation}
The necessary condition for any transmission is
\begin{equation}
\frac{{\mathbf u}_y \cdot {\mathbf k}_{TE}}{\mu_x} < 0
\Longleftrightarrow \frac{k_y}{\mu_x} < 0. \label{eq:physcond2}
\end{equation}
Clearly, the phase velocity is directed oppositely to the power flow
provided that $\mu_x < 0$. Negative refraction of the Poynting
vector requires that
\begin{equation}
k_x {\mathbf u}_x \cdot {\mathbf S}_{TE} < 0 \Longleftrightarrow
\mu_y < 0. \label{eq:negref}
\end{equation}

For the interface problem, we will also need to know conditions
for the existence of surface waves. The input impedance on the
surface filled by a uniaxial material is, for TE-polarized fields,
\begin{equation}
Z_{\rm inp} = {\omega \mu_x\over{\beta_{TE}}},
\end{equation}
where the normal component of the propagation factor reads
\begin{equation}
\beta_{TE} =\sqrt{{\mu_x\over{\mu_y}}\left(\omega^2\epsilon_z\mu_y
- k_x^2 \right)}
\end{equation}
with the square root branch defined by ${\rm Im}(\beta_{TE})<0$.
$k_x$ is the propagation factor along the surface. Thus, the
eigensolutions for an interface between this medium and free space
satisfy the following equation:
\begin{equation}
{\omega \mu_x \over {\beta_{TE}}}+{\omega \mu_0 \over {\beta_0}}=0
\end{equation}
where
\begin{equation}
\beta_0 =\sqrt{\omega^2\epsilon_0\mu_0 - k_x^2},
\end{equation}
with ${\rm Im}(\beta_0)<0$.

Let us consider now conditions for the existence of surface waves
along this interface. In this case both betas must be imaginary,
of course with negative imaginary parts:
$\beta_{TE}=-j\alpha_{TE}$ and $\beta_0=-j\alpha_0$, where
$\alpha_{TE}>0$ and $\alpha_0>0$. The eigenvalue equation becomes
\begin{equation}
{\omega \mu_x\over{\alpha_{TE}}}+{\omega \mu_0\over{\alpha_0}}=0.
\label{surface}
\end{equation}
Obviously, if all the material parameters are positive, this
equation has no solutions, but if $\mu_x<0$, surface wave
solutions are possible. This is well known for interfaces with
free-electron plasma.

Five different cases will be considered in the following. In the
first case, we choose $\mu_x > 0$, $\mu_y <0$, and $\epsilon_z <
0$. The wave vector surface is a two-sheeted hyperboloid with axis
parallel to the $x$-axis. The asymptotes of the hyperbola in
$xy$-plane divide the plane into regions of complex and real wave
vectors. Waves that are propagating parallel to $y$-axis are
supposed to decay exponentially inside the slab.

In the second case, $\mu_x> 0$, $\mu_y < 0$, $\epsilon_z > 0$. The
wave vector surface is a two-sheeted hyperboloid with the axis
parallel to the surface normal. In our 2D-case, real wave vectors
exist inside the region bounded by the asymptotes of the hyperbola
associated to the wave vector surfaces.

As a third case, we consider the situation complementary to the
second case in the sense that the signs of all the parameters are
changed. Notice that this does not affect the shape of the wave
vector curves. We have $\mu_x < 0$, $\mu_y > 0$, and $\epsilon_z
<0$. In fact, the Poynting vector is refracted positively in this
case (see (\ref{eq:negref})), but this is an interesting case
anyway because of the aforementioned contrast with respect to the
second case.

The fourth case consists of the usual isotropic BW slab ($\mu_x =
\mu_y < 0$, and $\epsilon_z < 0$) where focusing and negative
refraction phenomena are present. The wave vector curves are
ellipses (in our case of two equal parameters they are circles),
and real wave vectors exist everywhere inside the slab.

The fifth case is specially chosen to show the existence of
surface waves in the case when $\mu_x < 0$, $\mu_y < 0$, and
$\epsilon_z > 0$. We can easily see from (\ref{eq:ksurf}) that
there are neither real wave vectors nor backward waves. However,
we have found that surface waves on the interface are easily
excited in this case. Let us now present the numerical results.

\section{Numerical Results and Comparison with the Theory}

In the first four cases, we show the electric field distribution
at three suitable chosen increasing time steps to illustrate the
wave propagation and refraction phenomena. In the fifth case, we
illuminate a rectangular cylinder to see the surface waves.
Whenever a constitutive parameter is said to be positive, it is
supposed to be a constant and equal to the free space permittivity
or permeability. Negative material parameters obey the Lorentzian
dispersion rule and equal $-\epsilon_0$ or $-\mu_0$ at the center
frequency.

\subsection{Case I: $\mu_x > 0$, $\mu_y < 0$, $\epsilon_z < 0$}

\begin{widetext}

\begin{figure}[htb]
\mbox{\subfigure[]{\epsfig{figure=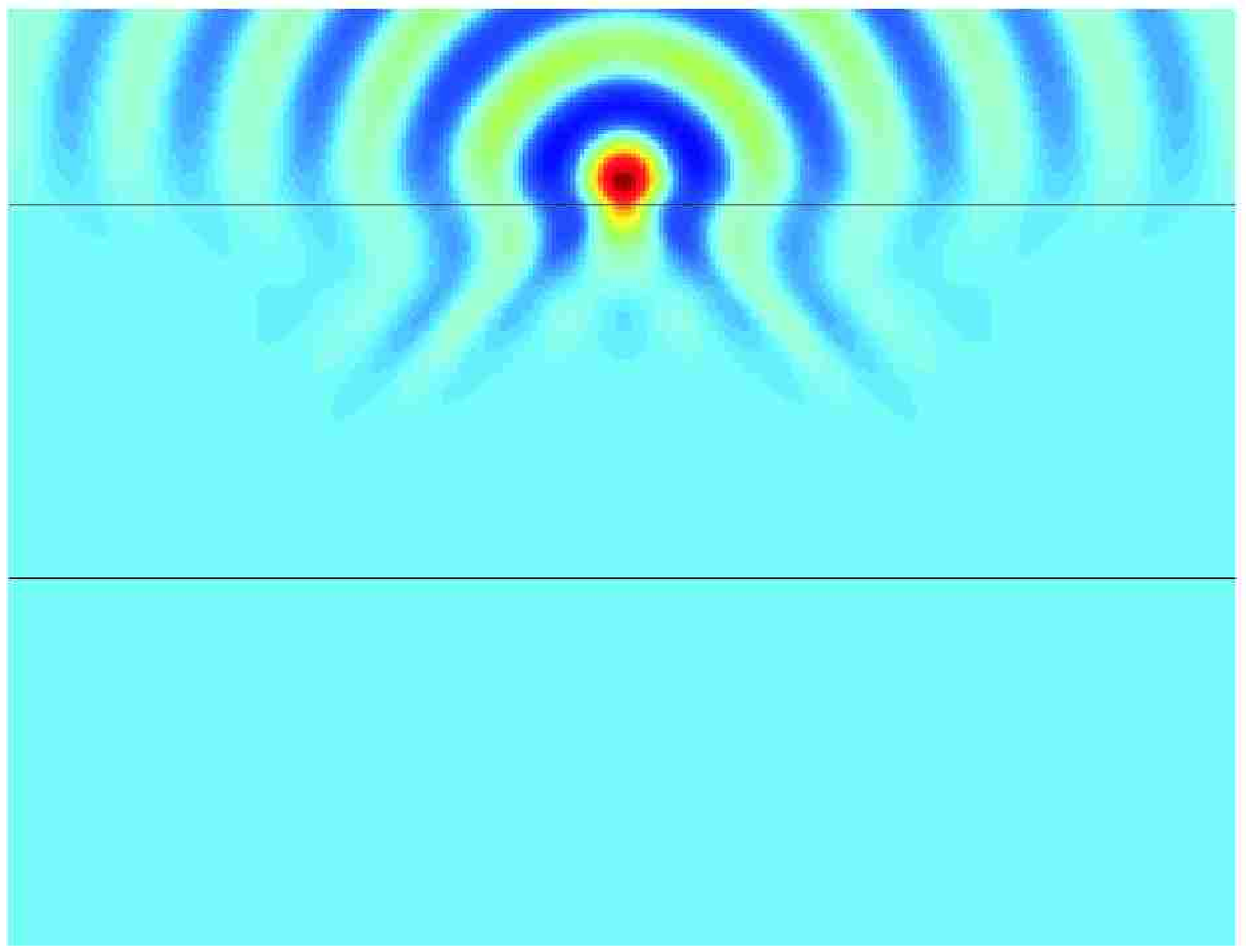,width=.32\textwidth,angle=0}}
\quad
\subfigure[]{\epsfig{figure=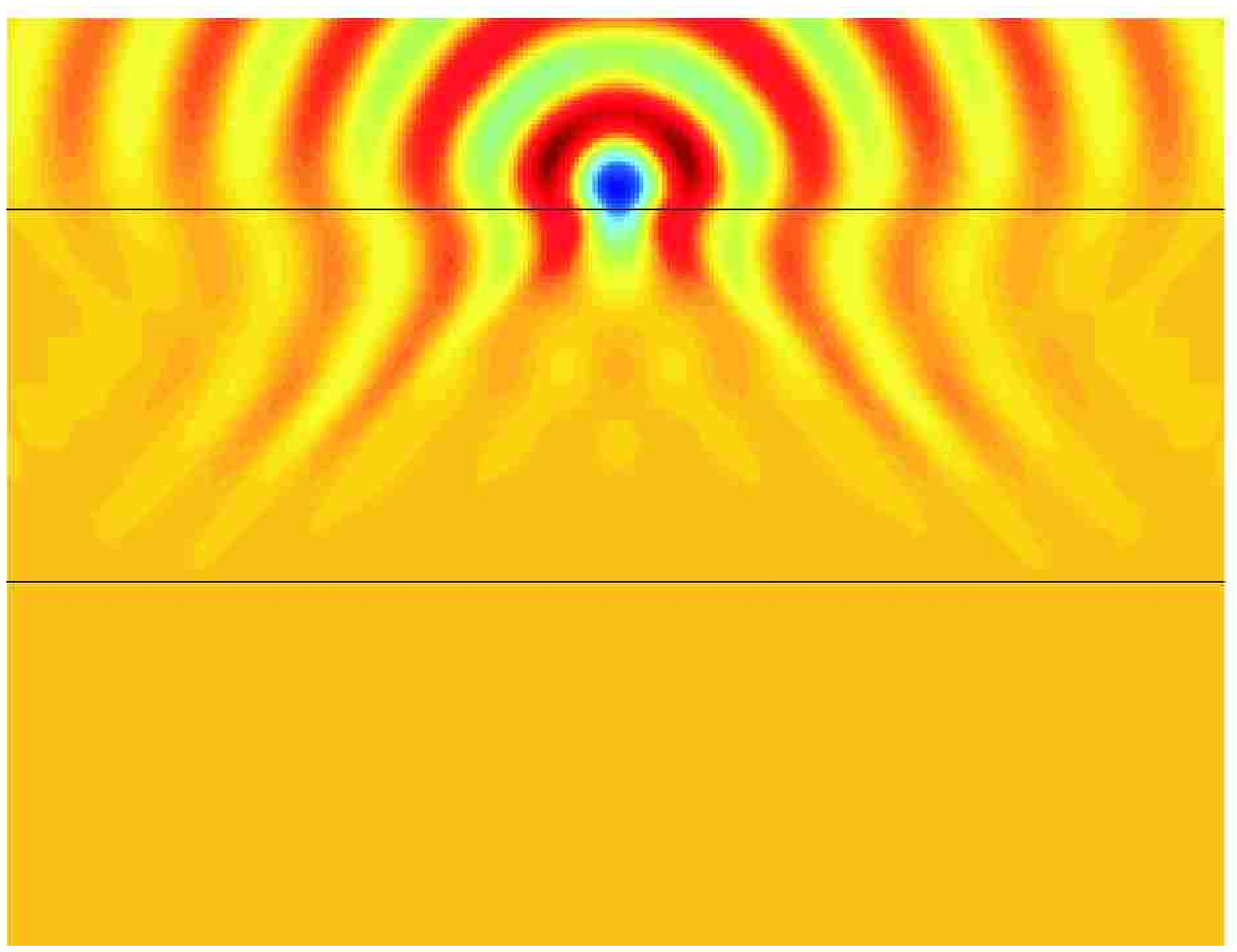,width=.32\textwidth,angle=0}}
\quad
\subfigure[]{\epsfig{figure=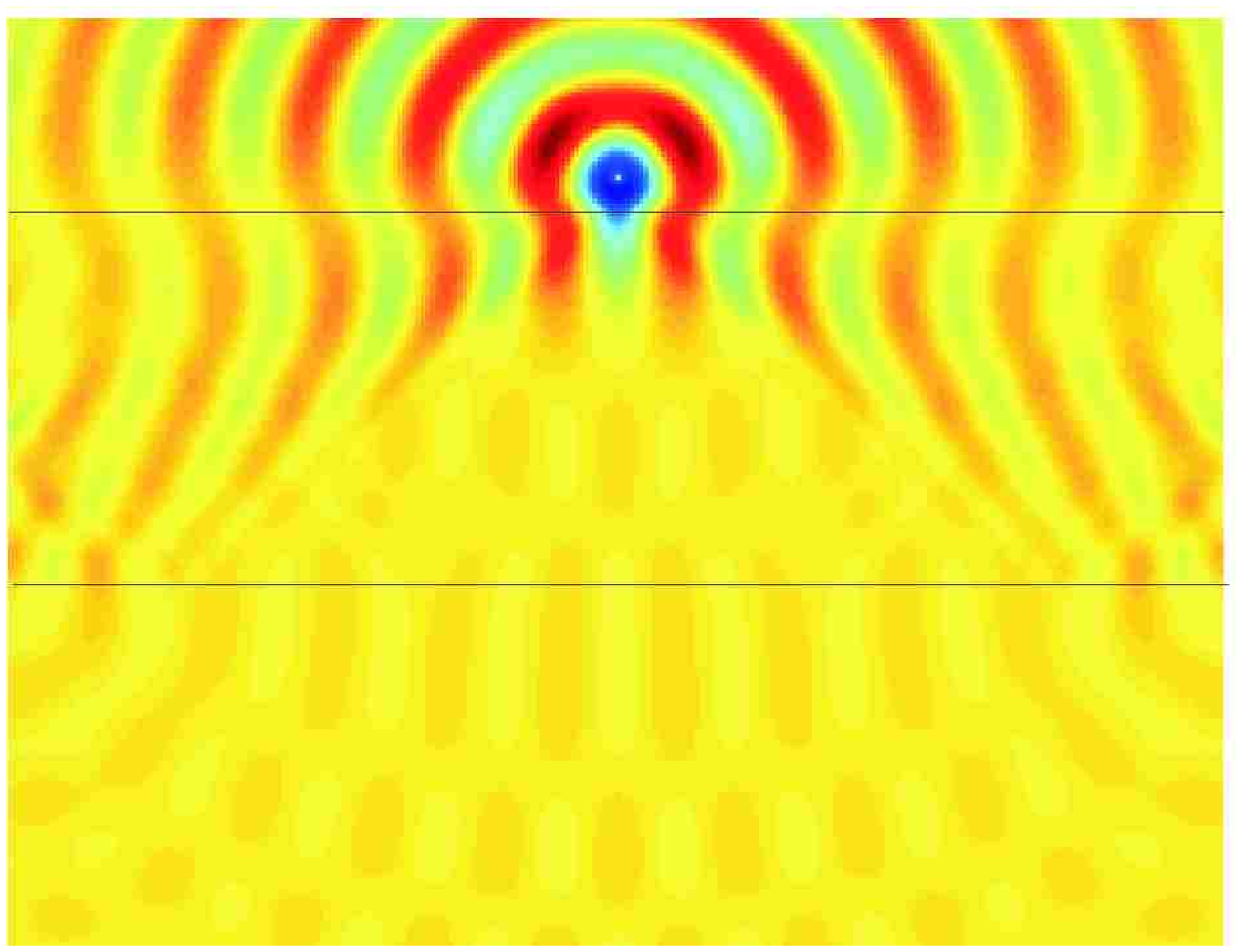,width=.32\textwidth,angle=0}}}
\caption{\small (a) The electric field $E_z$ penetrating into the
slab. Slab boundaries are indicated by black lines. (b) The
electric field within the slab. Notice the region in the center of
the slab, where the field amplitudes are very small. (c) It
appears that the wavefronts inside the slab are hyperbolas, in
agreement with the theory.} \label{kuva2}
\end{figure}

\end{widetext}

In this case, the theory shows that the wave vectors are complex
inside the slab within a region bounded by the asymptotes of a
hyperbola. Inspection of Figure \ref{kuva2} c) reveals that there
is indeed a region in the slab, where the electric field in
negligible all the time. There are some fields within the slab
just under the source. The hyperbola-shaped wavefronts propagate
obliquely downwards inside the slab and the power flow is
refracted negatively. The distance of the source from the first
interface is discretized with $10$ cells, and the thickness of the
slab corresponds to $80$ cells.

\subsection{Case II: $\mu_x > 0$, $\mu_y < 0$, $\epsilon_z > 0$}

\begin{widetext}

\begin{figure}[htb]
\mbox{\subfigure[]{\epsfig{figure=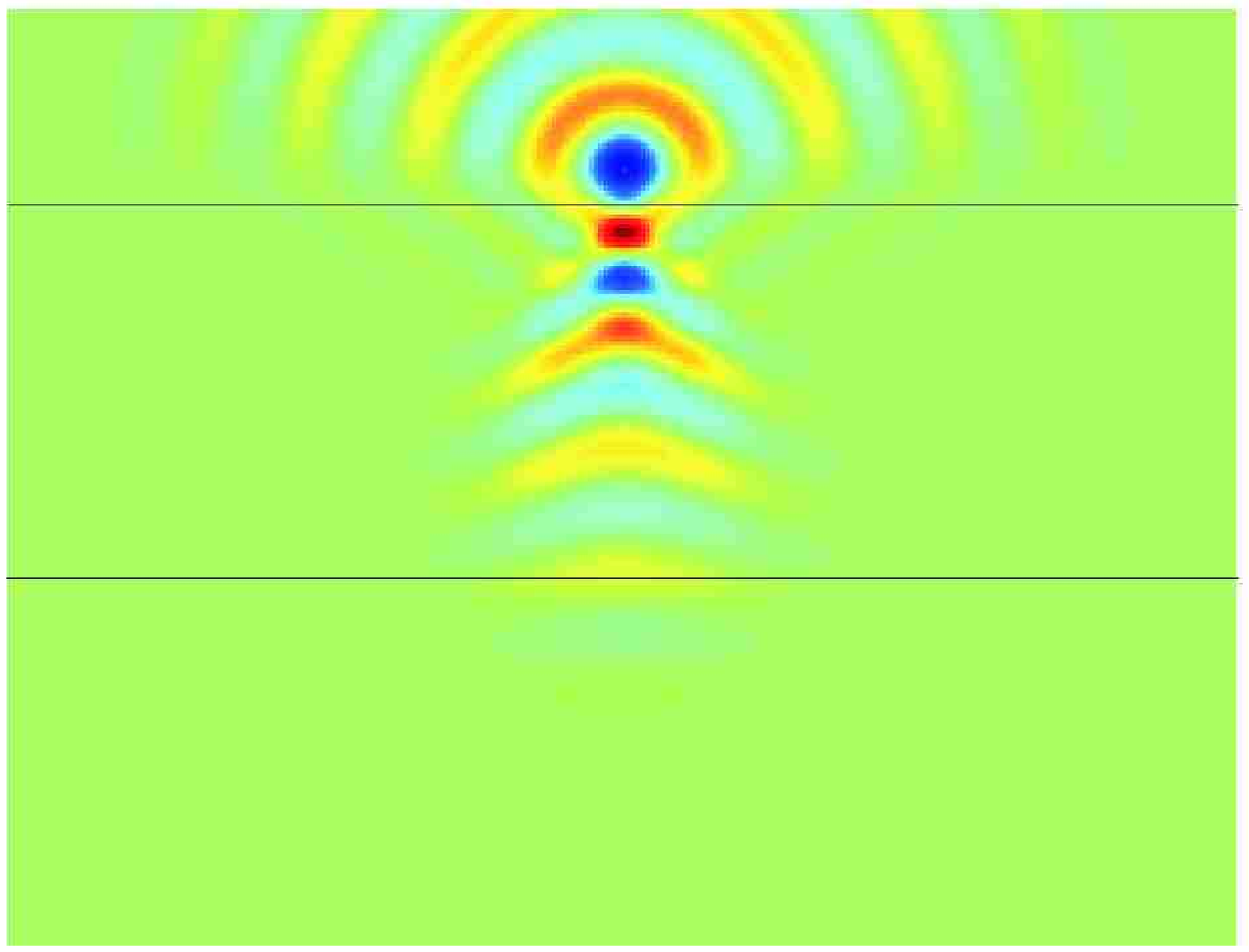,width=.32\textwidth,angle=0}}
\quad
\subfigure[]{\epsfig{figure=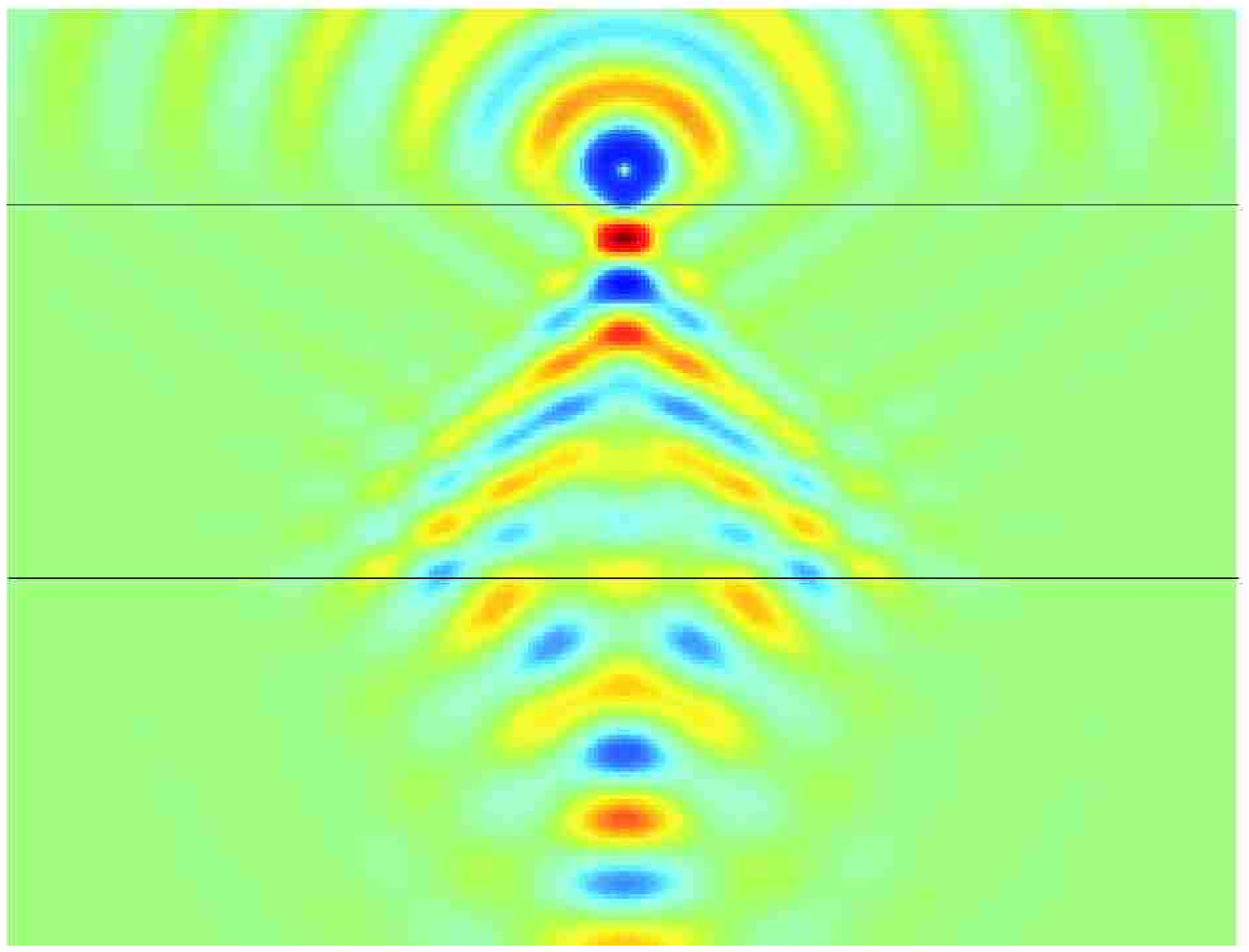,width=.32\textwidth,angle=0}}
\quad
\subfigure[]{\epsfig{figure=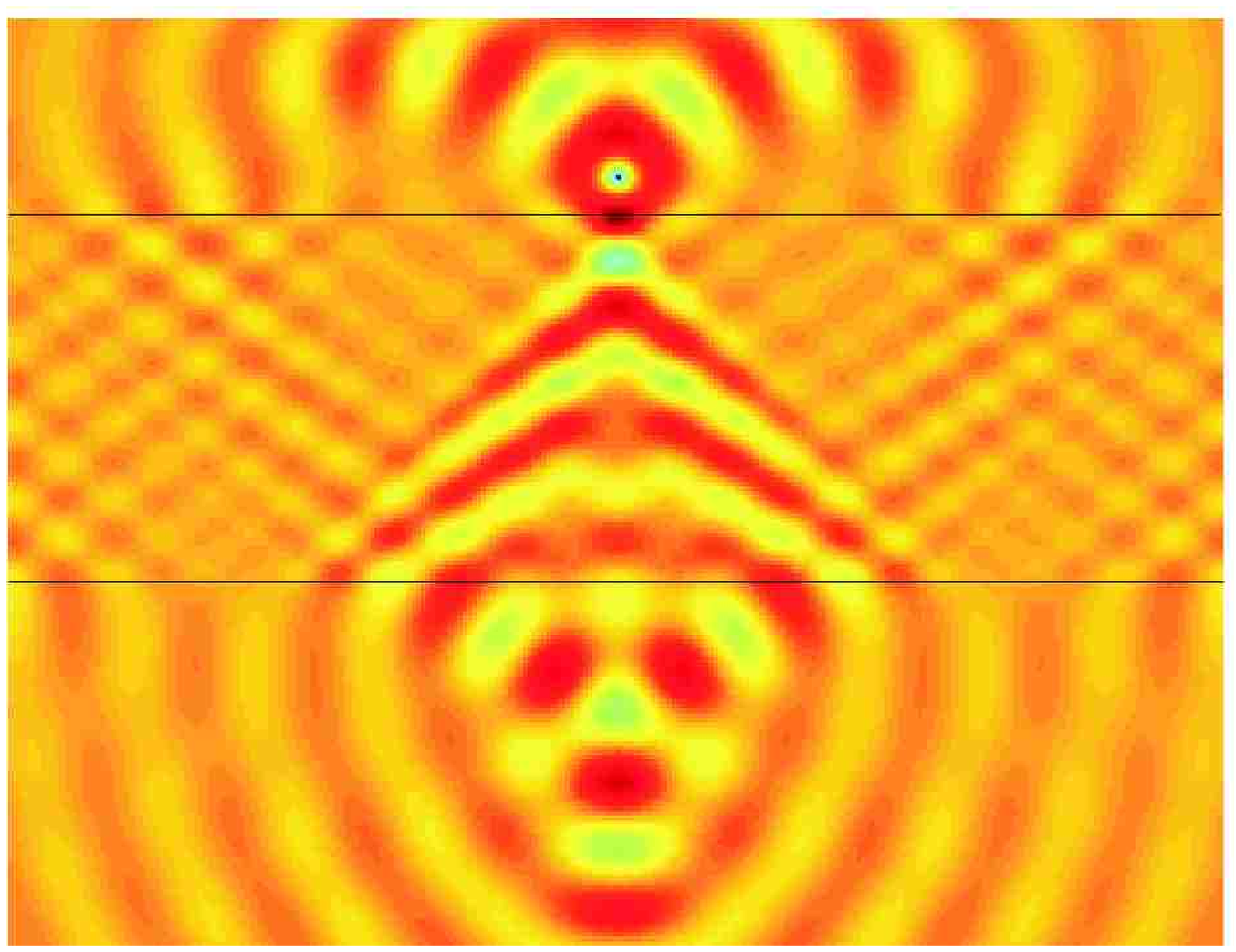,width=.32\textwidth,angle=0}}}
\caption{\small (a) The electric field $E_z$ penetrating into the
slab. (b) The electric field within the slab. Some waves have
already passed through the slab. It is seen that there are small
fields in the region, where the wave vector is complex, as
predicted by the theory. (c) Dispersive effects are more clearly
seen inside the slab. The electric field is concentrated to the
lower side of the slab. The wavefronts behind the slab outside the
region of large amplitudes are prolate ellipses. $d_s=10 \Delta y,
d=80 \Delta y$.} \label{kuva1}
\end{figure}

\end{widetext}

In Figure \ref{kuva1} a), a cylindrical wave is penetrating into
the BW slab. In Figure \ref{kuva1} b), some numerical dispersion
is visible. There are significant fields inside the region where
the theory yields real wave vectors, while the fields are rather
small elsewhere. Despite some dispersive effects, the wavefronts
of constant field value are reminiscent of hyperbolas. The phase
velocity is directed downwards. Some weak focusing of the power
flow is seen in Figure \ref{kuva1} b). The phase velocity inside
the slab is directed downwards, as can be seen from the theory.

\subsection{Case III: $\mu_x < 0$, $\mu_y > 0$, $\epsilon_z < 0$}

\begin{widetext}

\begin{figure}[htb]
\mbox{\subfigure[]{\epsfig{figure=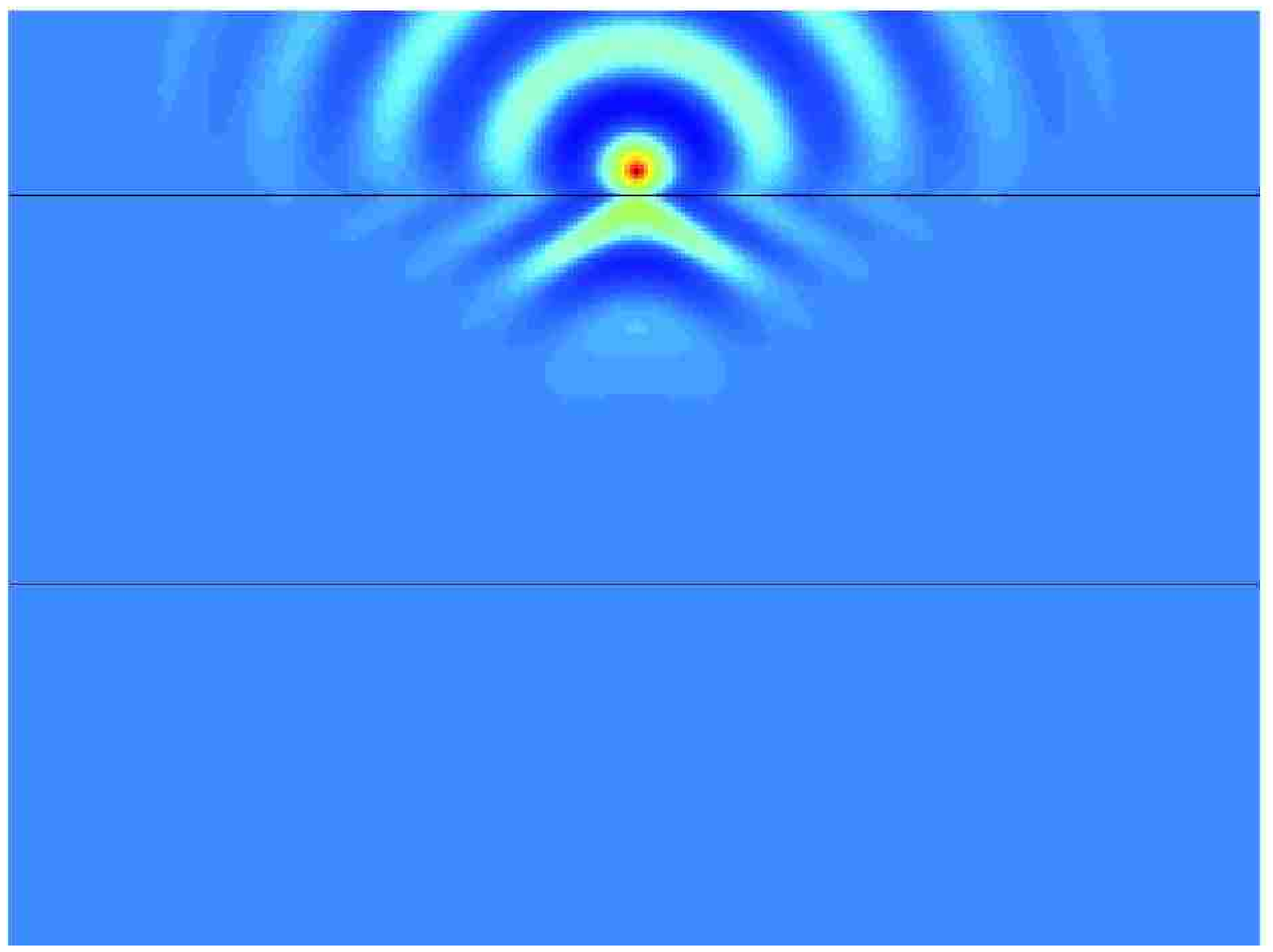,width=.32\textwidth,angle=0}}
\quad
\subfigure[]{\epsfig{figure=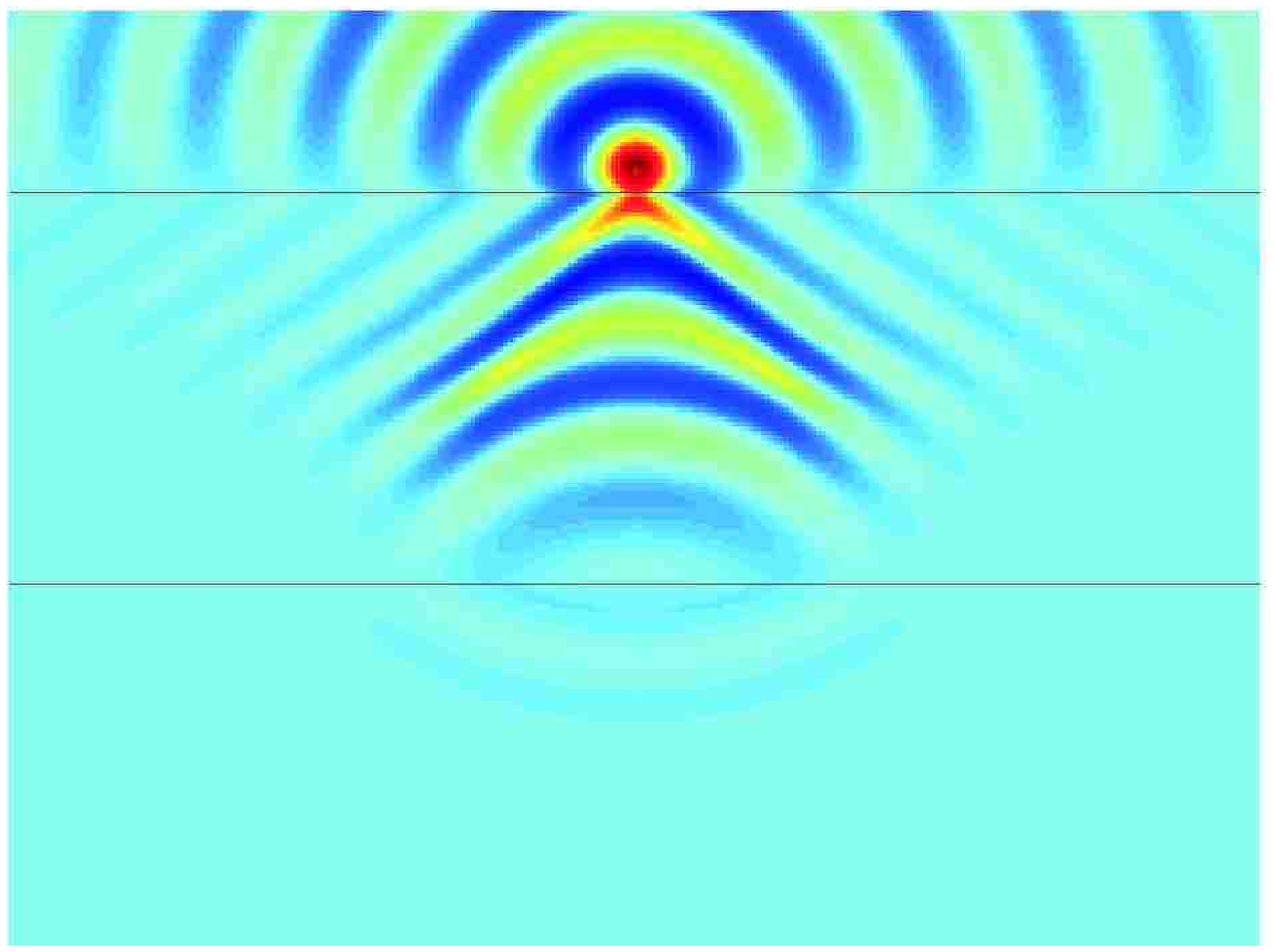,width=.32\textwidth,angle=0}}
\quad
\subfigure[]{\epsfig{figure=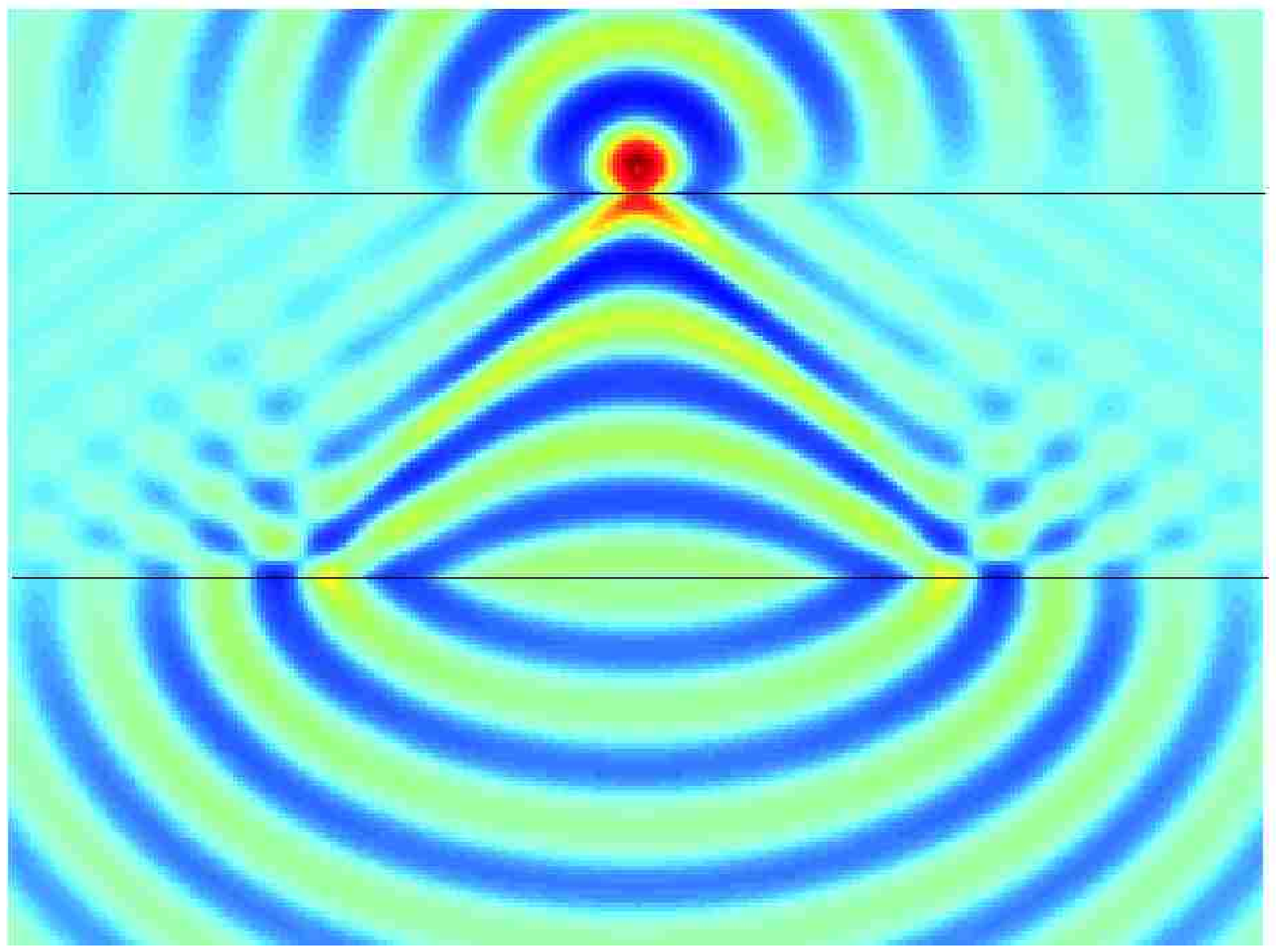,width=.32\textwidth,angle=0}}}
\caption{\small (a) The electric field $E_z$ penetrating into the
slab. Wavefronts are dramatically bent. (b) The electric field
within the slab. Some waves have already passed through the slab.
The wavefronts are seen to be hyperbolas. (c) Interestingly, the
wavefronts behind the slab appear to be ellipses with the center
on the lower interface of the slab. $d_s=10 \Delta y, d=80 \Delta
y$} \label{kuva3}
\end{figure}

\end{widetext}

To complete the analysis, we change the signs of the parameters of
the second case. Clearly, the wave vector surfaces as defined by
(\ref{eq:ksurf}) are not changed. In fact, the Poynting vector is
refracted positively in this case. However, the phase velocity is
directed upwards [see (\ref{eq:physcond2}) and (\ref{eq:negref})].
This phenomenon is clearly seen during the simulation. From Figure
\ref{kuva3} we see that the electric field distributions inside
the slab are quite similar to those of Figure \ref{kuva1} except
that the wavefronts are less distorted in Figure \ref{kuva3}. In
Figure \ref{kuva3} c), the wavefronts behind the slab are seen to
be oblate ellipses with the center on the lower interface of the
slab.

\subsection{Case IV: an Isotropic Slab with
$\mu_x < 0$, $\mu_y < 0$, $\epsilon_z < 0$}

\begin{widetext}

\begin{figure}[htb]
\mbox{\subfigure[]{\epsfig{figure=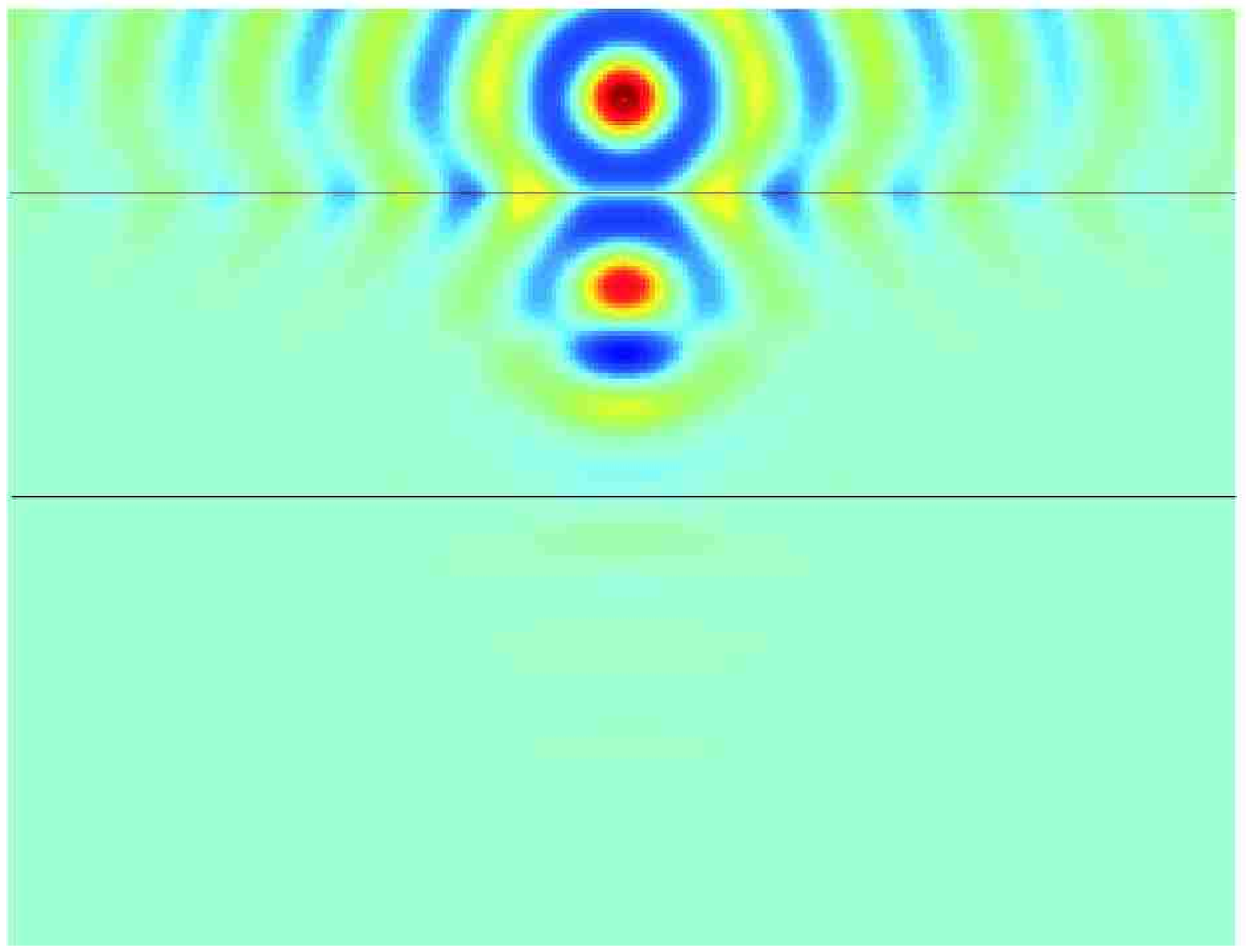,width=.32\textwidth,angle=0}}
\quad
\subfigure[]{\epsfig{figure=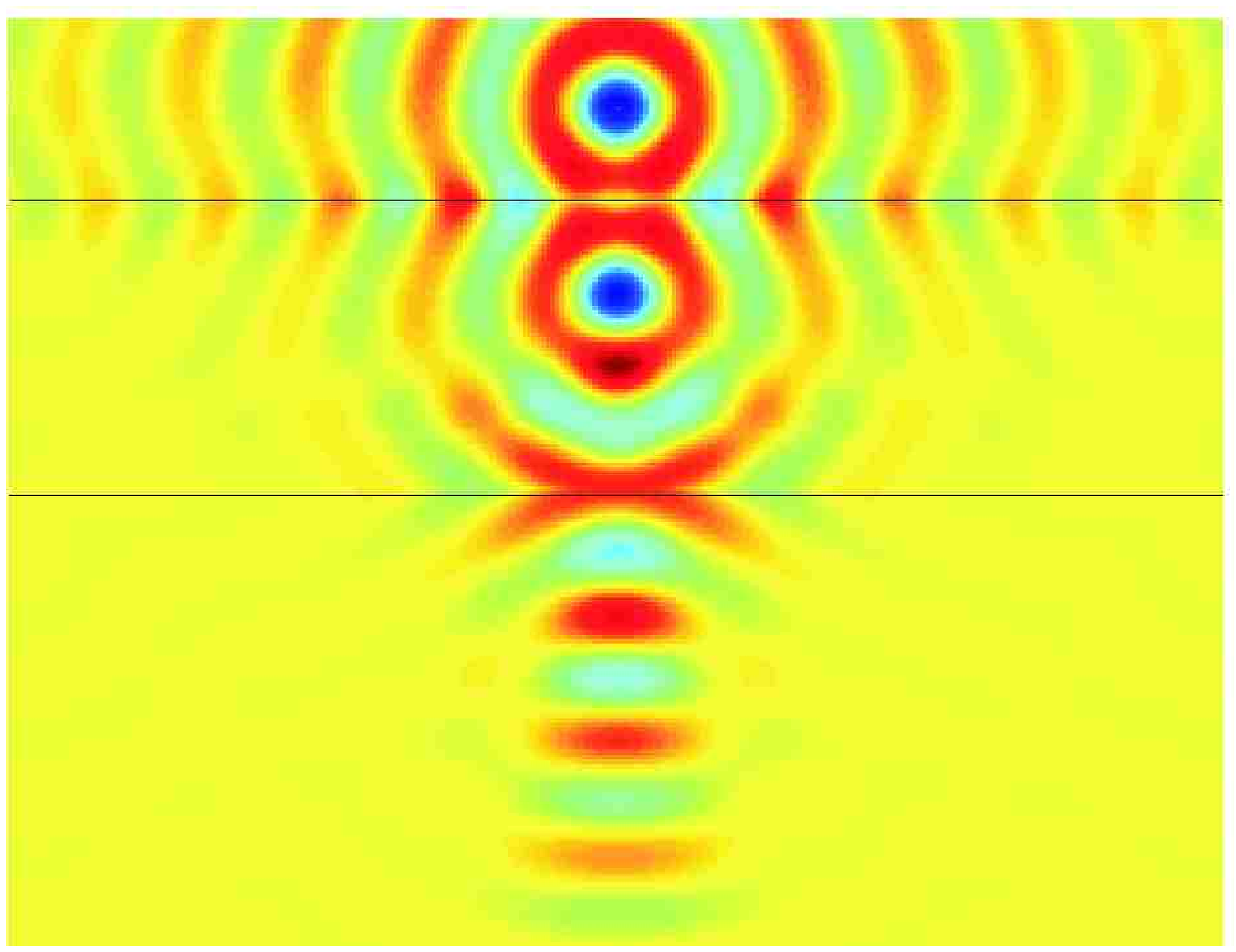,width=.32\textwidth,angle=0}}
\quad
\subfigure[]{\epsfig{figure=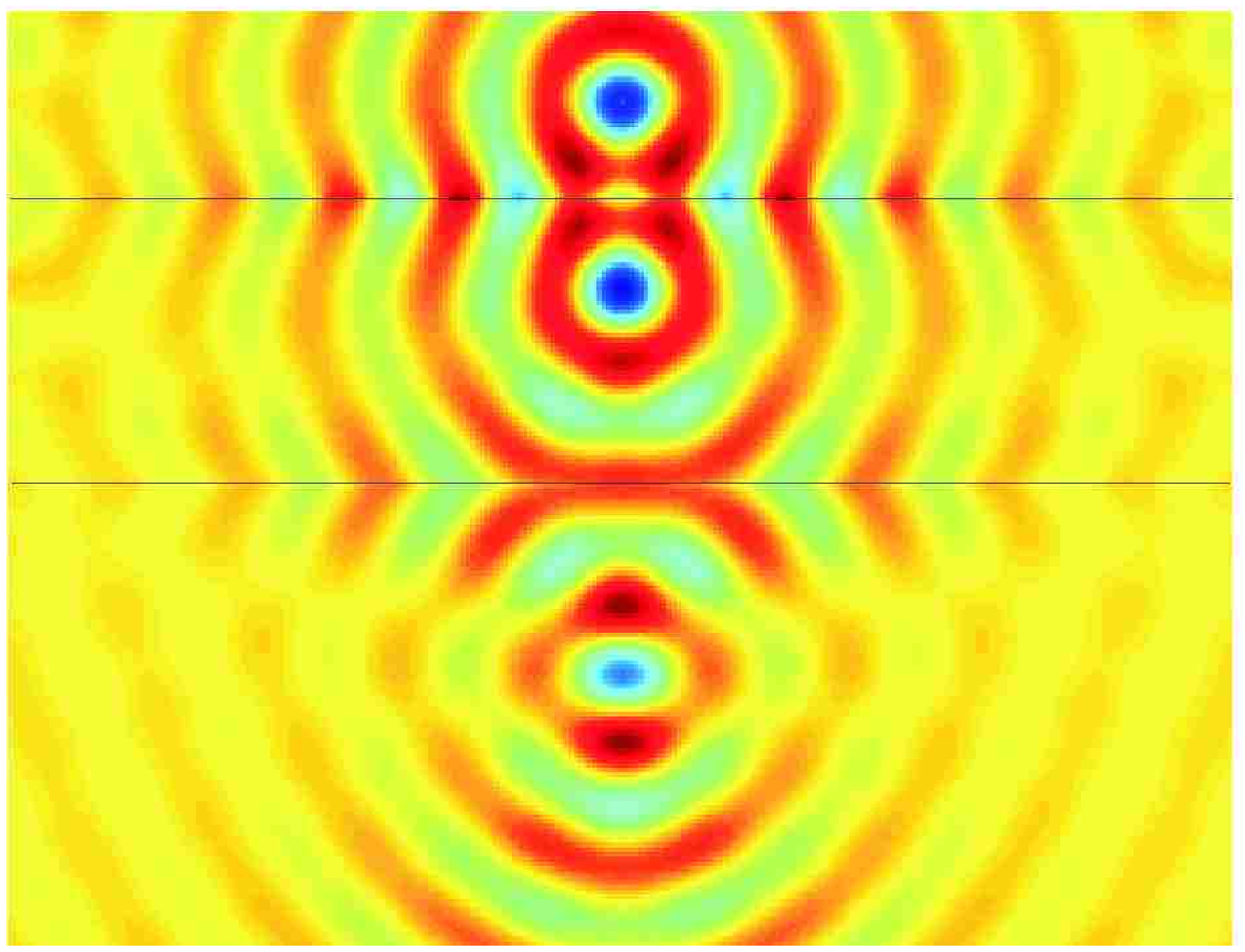,width=.32\textwidth,angle=0}}}
\caption{\small (a) The electric field $E_z$ penetrating into an
isotropic BW slab. (b) The electric field within the isotropic BW
slab. Some waves have already passed through the slab. The first
focus inside the slab is seen in the figure. (c) The second focus
behind the BW slab has become visible. $d_s=20 \Delta y, d=60
\Delta y$} \label{kuva4}
\end{figure}

\end{widetext}

Here we consider the usual isotropic BW (or double negative) slab
with all the relative material parameters close to minus one. This
case has been studied, for example, by Ziolkowski and Heyman in
\cite{Ziolkowski} in the case of Drude slabs. We obtained quite
similar results with our alternative discretization technique in
the case of isotropic Lorentz medium. The electric field
distributions are shown in Figure \ref{kuva4}. We can calculate
the positions of the foci from the slab thickness $d$ and the
distance $d_s$ of the source from the interface. Notice that we
must have $d_s < d$ to have a focus inside the slab. The foci
should appear at $y=-d_s$ inside the slab and at $y=-(2d-d_s)$
behind the slab. The appropriate derivations can be found in
\cite{Ziolkowski}.

From Figure \ref{kuva4} a) we see that the electric field is
concentrated in the expected position of the first focus. In
Figure \ref{kuva4} c), the second focus behind the slab is also
visible. It takes some time for the foci to develop, as is seen
from Figure \ref{kuva4} b), where the second focus is not yet
seen. The incident spectrum has some small components for which
the relative material parameters are not exactly minus unity.
Hence the wavefronts are not perfect circles as predicted by the
theory. Anyway, these results are in agreement with the theory
concerning negative refraction of the Poynting vector.
However, no ``perfect" focusing
has been observed, meaning that the focus area
is always not smaller that about half wavelength.
Steady state solutions for the
foci were not obtained. This same observation was also made by
Ziolkowski and Heyman in \cite{Ziolkowski}.

\subsection{Case V: $\mu_x < 0$, $\mu_y < 0$, $\epsilon_z > 0$}

\begin{figure}[htb]
\mbox{\subfigure[]{\epsfig{figure=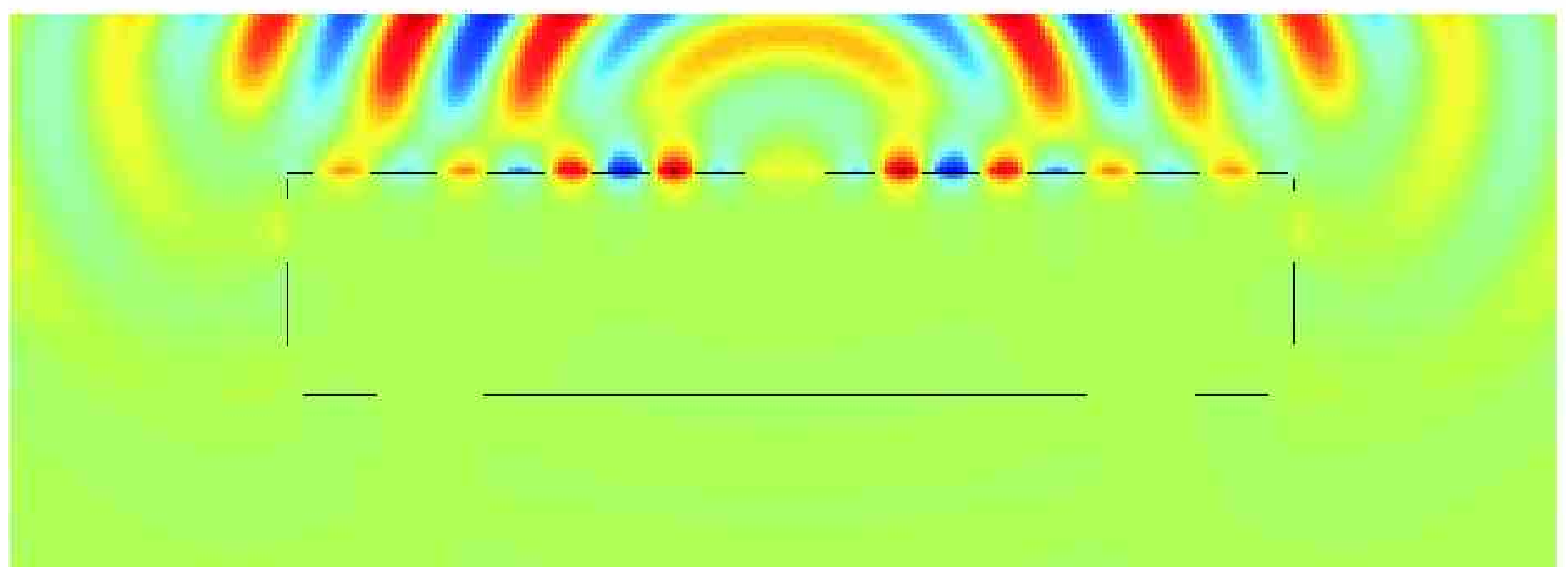,width=.23\textwidth,angle=0}}
\quad
\subfigure[]{\epsfig{figure=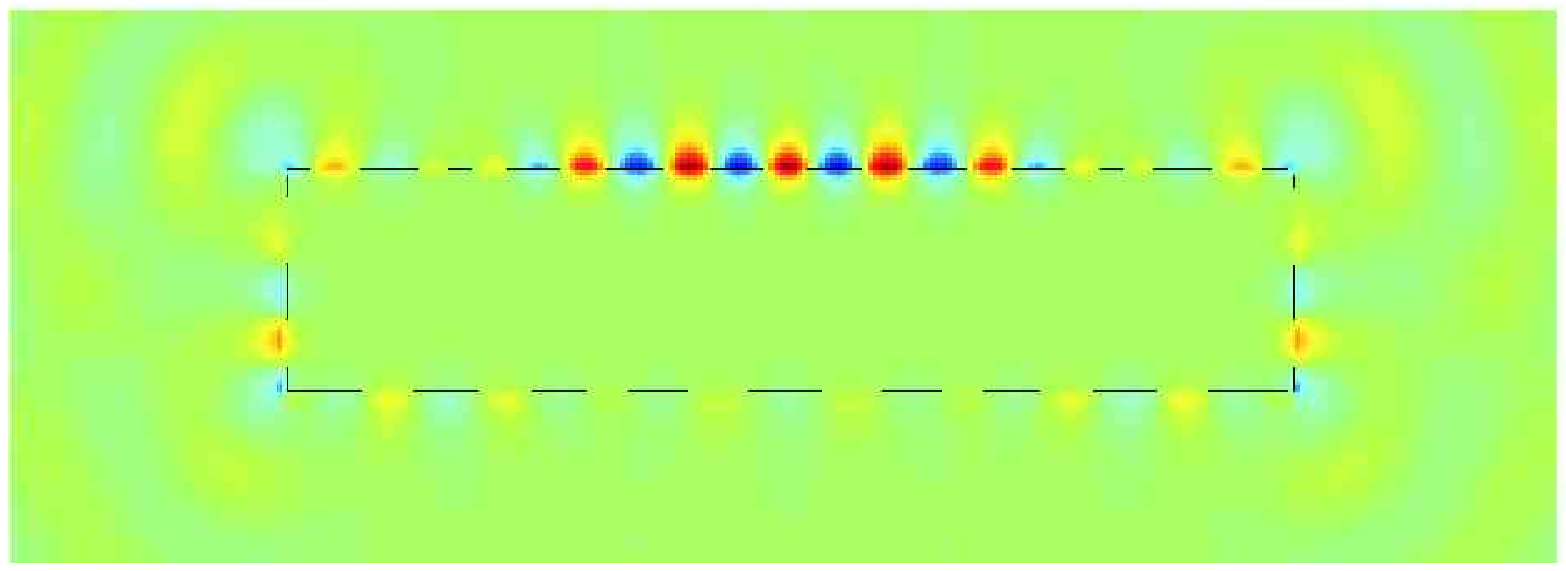,width=.23\textwidth,angle=0}}}
\caption{\small (a) The surface waves begin to develop. The source
is still on. (b) The source has been switched off, and surface
waves exist around the cylinder. As is known from the theory,
there are negligible fields inside the cylinder.} \label{kuva5}
\end{figure}

For this set of material parameters, the theory predicts that the
waves decay exponentially everywhere inside the slab. However, we
have found that surface waves are easily excited in this case, in
accordance with  the theoretical prediction, see \r{surface}. To
see this phenomenon, we illuminate a rectangular cylinder with a
pulse having a slightly broader spectrum. The electric field
distribution induced on the surface of the cylinder is shown in
Figure \ref{kuva5}.
Figure \ref{kuva5} a) shows how the surface waves begin to
develop. The source is still on. In Figure \ref{kuva5} b), the
source above the cylinder has been switched off. The surface waves
have propagated to the opposite side of the cylinder as well. In
agreement with the theory, there are no fields inside the cylinder
except for the immediate vicinity of the surface of the cylinder.

\section{Conclusions}

Wave propagation and refraction phenomena in uniaxially
anisotropic BW slabs have been studied with the FDTD method.
Special attention was paid to the shape of the wavefronts and to
the regions inside the slabs, where the wave vector becomes
complex, thus resulting in exponentially decaying waves. The
numerical results for anisotropic BW slabs were seen to
qualitatively agree well with the theoretical results. The effects
of negative refraction, (imperfect) focusing, and surface wave
excitation have been demonstrated. Potentially useful
transformations of wave fronts between spherical, elliptical, and
hyperbolical can be realized in homogeneous uniaxial backward wave
slabs.

\section*{Acknowledgments}

Financial support was received from Nokia Foundation and Emil
Aaltonen Foundation. Useful discussions with S.A. Tretyakov, S.I.
Maslovski and P.A. Belov are gratefully acknowledged.

\bibliography{metapap}

\begin{thebibliography}{11}
\expandafter\ifx\csname natexlab\endcsname\relax\def\natexlab#1{#1}\fi
\expandafter\ifx\csname bibnamefont\endcsname\relax
  \def\bibnamefont#1{#1}\fi
\expandafter\ifx\csname bibfnamefont\endcsname\relax
  \def\bibfnamefont#1{#1}\fi
\expandafter\ifx\csname citenamefont\endcsname\relax
  \def\citenamefont#1{#1}\fi
\expandafter\ifx\csname url\endcsname\relax
  \def\url#1{\texttt{#1}}\fi
\expandafter\ifx\csname urlprefix\endcsname\relax\def\urlprefix{URL }\fi
\providecommand{\bibinfo}[2]{#2}
\providecommand{\eprint}[2][]{\url{#2}}

\bibitem[{\citenamefont{Lindell et~al.}(2001)\citenamefont{Lindell, Tretyakov,
  Nikoskinen, and Ilvonen}}]{Lindell}
\bibinfo{author}{\bibfnamefont{I.}~\bibnamefont{Lindell}},
  \bibinfo{author}{\bibfnamefont{S.}~\bibnamefont{Tretyakov}},
  \bibinfo{author}{\bibfnamefont{K.}~\bibnamefont{Nikoskinen}},
  \bibnamefont{and} \bibinfo{author}{\bibfnamefont{S.}~\bibnamefont{Ilvonen}},
  \bibinfo{journal}{Microw. and Opt. Tech. Lett.}
  \textbf{\bibinfo{volume}{31}}, \bibinfo{pages}{129} (\bibinfo{year}{2001}).

\bibitem[{\citenamefont{Veselago}(1968)}]{Veselago}
\bibinfo{author}{\bibfnamefont{V.}~\bibnamefont{Veselago}},
  \bibinfo{journal}{Sov. Phys. Uspekhi} \textbf{\bibinfo{volume}{10}},
  \bibinfo{pages}{509} (\bibinfo{year}{1968}).

\bibitem[{\citenamefont{Valanju and Walser}(2002)}]{Valanju}
\bibinfo{author}{\bibfnamefont{P.}~\bibnamefont{Valanju}} \bibnamefont{and}
  \bibinfo{author}{\bibfnamefont{R.}~\bibnamefont{Walser}},
  \bibinfo{journal}{Phys. Rev. Lett.} \textbf{\bibinfo{volume}{88}},
  \bibinfo{pages}{187401} (\bibinfo{year}{2002}).

\bibitem[{\citenamefont{Pendry}(2000)}]{Pendry}
\bibinfo{author}{\bibfnamefont{J.}~\bibnamefont{Pendry}},
  \bibinfo{journal}{Phys. Rev. Lett.} \textbf{\bibinfo{volume}{88}},
  \bibinfo{pages}{3966} (\bibinfo{year}{2000}).

\bibitem[{\citenamefont{Ziolkowski and Heyman}(2001)}]{Ziolkowski}
\bibinfo{author}{\bibfnamefont{R.}~\bibnamefont{Ziolkowski}} \bibnamefont{and}
  \bibinfo{author}{\bibfnamefont{E.}~\bibnamefont{Heyman}},
  \bibinfo{journal}{Phys. Rev. E} \textbf{\bibinfo{volume}{53}},
  \bibinfo{pages}{135} (\bibinfo{year}{2001}).

\bibitem[{\citenamefont{Lindell and Ilvonen}(2002)}]{Lindell2}
\bibinfo{author}{\bibfnamefont{I.}~\bibnamefont{Lindell}} \bibnamefont{and}
  \bibinfo{author}{\bibfnamefont{S.}~\bibnamefont{Ilvonen}},
  \bibinfo{journal}{J. of Electromagn. Waves and Appl.}
  \textbf{\bibinfo{volume}{16}}, \bibinfo{pages}{303} (\bibinfo{year}{2002}).

\bibitem[{\citenamefont{Tretyakov et~al.}(2002)\citenamefont{Tretyakov,
  Nefedov, Simovski, and Maslovski}}]{Tretyakov}
\bibinfo{author}{\bibfnamefont{S.}~\bibnamefont{Tretyakov}},
  \bibinfo{author}{\bibfnamefont{I.}~\bibnamefont{Nefedov}},
  \bibinfo{author}{\bibfnamefont{C.}~\bibnamefont{Simovski}}, \bibnamefont{and}
  \bibinfo{author}{\bibfnamefont{S.}~\bibnamefont{Maslovski}},
  \emph{\bibinfo{title}{Advances in Electromagnetics of Complex Media and
  Metamaterials}} (\bibinfo{publisher}{Kluwer}, \bibinfo{year}{2002}).

\bibitem[{\citenamefont{Smith et~al.}(2000)\citenamefont{Smith, Padilla, Vier,
  Nemat-Nasser, and Schultz}}]{Smith}
\bibinfo{author}{\bibfnamefont{D.}~\bibnamefont{Smith}},
  \bibinfo{author}{\bibfnamefont{W.}~\bibnamefont{Padilla}},
  \bibinfo{author}{\bibfnamefont{D.}~\bibnamefont{Vier}},
  \bibinfo{author}{\bibfnamefont{S.}~\bibnamefont{Nemat-Nasser}},
  \bibnamefont{and} \bibinfo{author}{\bibfnamefont{S.}~\bibnamefont{Schultz}},
  \bibinfo{journal}{Phys. Rev. Lett.} \textbf{\bibinfo{volume}{84}},
  \bibinfo{pages}{4184} (\bibinfo{year}{2000}).

\bibitem[{\citenamefont{Taflove and Hagness}(2000)}]{Taflove}
\bibinfo{author}{\bibfnamefont{A.}~\bibnamefont{Taflove}} \bibnamefont{and}
  \bibinfo{author}{\bibfnamefont{S.}~\bibnamefont{Hagness}},
  \emph{\bibinfo{title}{Computational Electrodynamics -- The finite-difference
  time-domain method}} (\bibinfo{publisher}{Artech House},
  \bibinfo{address}{Boston}, \bibinfo{year}{2000}).

\bibitem[{\citenamefont{Young and Nelson}(2001)}]{Young}
\bibinfo{author}{\bibfnamefont{J.}~\bibnamefont{Young}} \bibnamefont{and}
  \bibinfo{author}{\bibfnamefont{R.}~\bibnamefont{Nelson}},
  \bibinfo{journal}{IEEE Antennas Propag. Magazine}
  \textbf{\bibinfo{volume}{43}}, \bibinfo{pages}{72} (\bibinfo{year}{2001}).

\bibitem[{\citenamefont{K$\ddot{a}$rkk$\ddot{a}$inen and
  Maslovski}(2003)}]{motl2}
\bibinfo{author}{\bibfnamefont{M.}~\bibnamefont{K$\ddot{a}$rkk$\ddot{a}$inen}}
  \bibnamefont{and}
  \bibinfo{author}{\bibfnamefont{S.}~\bibnamefont{Maslovski}},
  \bibinfo{journal}{Microw. and Opt. Tech. Lett.} \textbf{\bibinfo{volume}{5}}
  (\bibinfo{year}{2003}).

\end{thebibliography}

\end{document}